\documentclass[aps,pra,reprint]{revtex4-2}

\usepackage{amsmath,amssymb}
\usepackage{graphicx}
\usepackage{dcolumn}
\usepackage{bm}

\usepackage[colorlinks=true, allcolors=blue]{hyperref}
\usepackage{physics}
\usepackage[caption=false]{subfig}

\usepackage{xcolor}
\usepackage{soul}

\begin{document}

\title{Identification and properties of topological states in the bulk of quasicrystals}

\author{Frode Balling-Ansø}
 \affiliation{Department of Physics and Astronomy, Aarhus University, DK-8000 Aarhus C, Denmark}
 \author{Jeppe Lykke Krogh}
 \affiliation{Department of Physics and Astronomy, Aarhus University, DK-8000 Aarhus C, Denmark}
  \author{Ella Elisabeth Lassen}
 \affiliation{Department of Physics and Astronomy, Aarhus University, DK-8000 Aarhus C, Denmark}
\author{Anne E. B. Nielsen}
\affiliation{Department of Physics and Astronomy, Aarhus University, DK-8000 Aarhus C, Denmark}

\begin{abstract}
In contrast to the usual bulk-boundary correspondence, topological states localized within the bulk of the system have been numerically identified in quasicrystalline structures, termed bulk localized transport (BLT) states. These states exhibit properties different from edge states, one example being that the number of BLT states scales with system size, while the number of edge states scales with system perimeter. Here, we define a procedure to identify BLT states, which is based on the physically motivated crosshair marker and robustness analyses. Applying the procedure to the Hofstadter model on the Ammann-Beenker tiling, we find that the BLT states appear mainly for magnetic fluxes within a specific interval. While edge states appear at low densities of states, we find that BLT states can appear at many different densities of states. Many of the BLT states are found to have real-space localization that follows geometric patterns characteristic of the given quasicrystal. Furthermore, BLT states can appear both isolated and in groups within the energy spectrum which could imply greater robustness for the states within such groups. The spatial localization of the states within a certain group can change depending on the Fermi energy. 
\end{abstract}

\maketitle

\section{Introduction}

During the past decades, the theory of topological systems \cite{top4, top1, top3, top2} has become an increasingly prevalent way of understanding phases of matter beyond Ginzburg-Landau theory \cite{Landau} and renormalization group \cite{RG}. Here, phases of matter are described and distinguished using topological constants, such as the Chern number, which are invariant under adiabatic deformation. At the heart of this theory lies the concept of bulk-boundary correspondence \cite{BBtheorem}. This dictates that a system described by a non-trivial topological constant in the thermodynamic limit must have conducting edge states when it has open boundary conditions. This is all well-understood for the usual cases of crystalline lattice structures described by a unit cell. 

Recently, the study of topological insulators has expanded to graphs that cannot be described by a unit cell such as self-similar structures \cite{selfsim1, selfsim2, selfsim3, Saswat} that stretch the idea of what can be considered a boundary. Another case is that of topology in quasicrystalline structures \cite{QS2, QS1}. One property of these structures is a lack of periodicity, but they still have a well-defined boundary and bulk. Nevertheless, some quasicrystals have been shown to contradict the standard bulk-boundary correspondence by hosting bulk localized transport (BLT) states which, unlike robust edge states, reside within the bulk of the system \cite{A1,A2}. In contrast we refer to topological edge states as edge localized transport (ELT) states. Studying BLT states is a worthwhile endeavor, as they behave in ways different from the usual edge states. As an example, the number of BLT states is expected to scale with system size while the number of ELT
states scales with the perimeter of the system. 

Describing and identifying these BLT states is not straightforward. In crystalline systems, the topology is seen based on topological invariants that become exact and robust integers in the thermodynamic limit. In finite systems, the topology is imperfect, an example being a small hybridization between edge states on opposite ends of a finite cylinder. The imperfect topology can, however, still be utilized experimentally if it is close enough to the ideal limit. For BLT states, there is generally not a limit, in which the topology becomes exact. As an example, two BLT states with opposite chiralities may stay at the same spatial distance from one another, even if the quasicrystal is made larger. The relevant question is hence how close a given state is to showing ideal topology with respect to what is needed for the intended purpose. This makes it harder to define exactly what constitutes a BLT state.  

BLT states have been analyzed \cite{A1} by applying the Bott index \cite{Bott} in quasicrystalline structures. This marker yields an integer which allows one to unambiguously distinguish between what is topological or not. However, it is not based on simple, physical quantities. As such, it is not clear how one might measure the Bott index in a real system.
Furthermore, the Bott index is computed by mapping the system onto a torus. The mapping introduces ambiguity in the definition of the marker. This further raises the question of how relevant the identified topological non-triviality is due to its unclear intuition. Another analysis \cite{A2} has been carried out using the local Chern marker \cite{Chernmarker}. This marker does not necessarily take on integer values. It is derived from a reformulation of the formula for the Chern number for systems with translational invariance, and its interpretation for systems that are far from periodic is hence not clear. 

We remedy these shortcomings by applying the crosshair marker \cite{crosshair}. This marker is derived without the restriction of translational invariance as a site-wise measure of conductance under an applied electric field and therefore holds a clear physical interpretation. This opens possibilities for experimental applications, and whenever the cross-hair marker identifies topological non-triviality we know in which way this topology manifests. This is important as the imperfect topology exhibited by the system demands a thorough understanding of what leads something to be classified as being topological. The marker also indicates where the system turns conducting and thus the localization of the induced current which allows us to distinguish between BLT and ELT states. Using this marker, we create a step-by-step procedure that can identify BLT
and ELT states in non-periodic systems. The procedure involves parameters that determine how close to topological a state needs to be to be flagged as topological by the procedure. The procedure applies two sorting steps: one for identifying possible topological states and one for measuring robustness and radial localization. 

Through application of this procedure to the Hofstadter model on the Ammann-Beenker tiling, we study the properties and characteristics of the topological states. We find that the number of BLT states scales with the number of sites in the system. This is in contrast to the number of edge states which scales with the perimeter of the system. This further enforces the idea that BLT states are different from topological edge states. We also find that whenever the magnetic flux is within a certain interval, the number of BLT states is significantly larger than when the flux lies outside this interval. Furthermore, in contrast to topological edge states, we find that BLT states can appear at any density of states. This could imply that other factors play a role towards topologically protecting these kinds of states. To this end, we look at the geometrical patterns exhibited by the BLT states in real-space. This reveals a tendency towards certain patterns formed by the geometry of the quasicrystal. Finally, we test the robustness of these states showing that their real-space localization resembles that of close-lying non-topological states. As such, robustness is favored whenever the states lie within groups of neighboring BLT states in the energy spectrum. Within such a group, the localization of the topological current may change depending on the Fermi energy of the system. 

The framework of the article is as follows: In Sec.\ \ref{sec:theory} we discuss the preliminary theory needed for the rest of the article. This includes introducing the Hamiltonian, its symmetries, and the methods we use for our identification scheme and analysis. In Sec.\ \ref{sec:alg} we define the procedure used to identify topological states in the system. In Sec.\ \ref{sec:data} we identify the number of topological states with variable external magnetic field and system size. In Sec.\ \ref{sec:BLT} we examine the characteristics of the BLT states. This includes graphing the density of states and comparing their tendencies with those of topological edge states. We investigate what geometric patterns BLT states exhibit and how they vary. We then discuss robustness by comparing them to close-lying non-topological states and studying how they are grouped in the energy spectrum. In Sec.\ \ref{sec:Discuss} we discuss any problems and possible amendments for how we choose to define BLT states. Finally, in Sec.\ \ref{sec:conclusion}, we summarize our results and elaborate on possible future developments.  

\section{Theoretical Framework} \label{sec:theory}

In this section, we introduce the theory needed to perform the numerical calculations. All calculations are done in natural units, $e = \hbar = 1$. In these units, the magnetic flux quantum is given by $\phi_0 = 2\pi$.

\subsection{Ammann-Beenker Tiling}

\begin{figure}
    \includegraphics[width=0.45\textwidth]{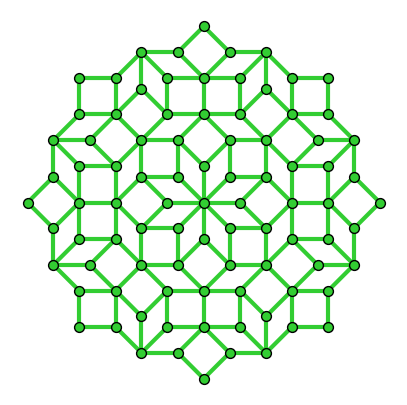}
    %\captionsetup{justification=centerlast} 
    \captionsetup{justification=justified}
    \caption{ Instance of Ammann-Beenker tiling with $89$ sites centered at the point of eightfold rotational symmetry. We choose a scale such that the distance between any two vertices connected by an edge is $1/\sqrt{2}$.}
    \label{fig:AB}
\end{figure}

The model we study is based on the quasicrystalline structure known as the Ammann-Beenker (AB) tiling \cite{Beenker, AB}. For certain boundary conditions this structure exhibits eight-fold rotational symmetry as in Fig.\ \ref{fig:AB}, but it has no translational symmetry. It is composed of two kinds of four-sided tiles, a square and a rhombus, which fit together in non-periodic ways. To generate the tiling, we apply the cut-and-project method \cite{AB}.

\subsection{The Hofstadter Model}

The model we consider is a single particle tight-binding Hamiltonian on the AB-tiling in the $xy$-plane. The system is exposed to a perpendicular magnetic field $\textbf{B} = B\hat{z}$, which enters the Hamiltonian as a Peierls phase $\theta_{ij}$ by
\begin{equation}\label{eq:Hamiltonian}
    H = - \sum_{\langle i,j \rangle} e^{i\theta_{ij}}\ket{\textbf{r}_j}\bra{\textbf{r}_i},
\end{equation}
where $\ket{\textbf{r}_j}$ is the state localized on a single site at \newline $\textbf{r}_j=(x_j, y_j)$ and $\langle i,j \rangle$ denotes pairs of connected sites in the graph as seen in Fig.\ \ref{fig:AB}. The Peierls phase is given by
\begin{equation}\label{eq:peierls}
    \theta_{ij} = \int_{\textbf{r}_i}^{\textbf{r}_j} \textbf{A}(\textbf{r}) \cdot d\textbf{r}, 
\end{equation}
where $\textbf{A}(\textbf{r})$ is the magnetic vector potential. This is known as the Hofstadter model \cite{Hofstadter}. Using the Landau gauge $\textbf{A} = Bx\hat{y}$ the phase can be calculated to be
\begin{equation}\label{eq:phase}
    \theta_{ij} = 2\pi \phi (y_j-y_i)(x_j+x_i),
\end{equation}
where $\phi$ is the flux through one of the squares of the AB-tiling given in units of the flux quantum $\phi_0$. Let $\ket{\psi_i}$ and $E_i$ denote the $i$th eigenstate of $H$ along with its eigenvalue indexed from the smallest to the highest energy, starting at $i=0$. 

If the graph is generated with eightfold rotational symmetry, this symmetry is carried over to the Hamiltonian model as a combined operation of a rotation and a gauge transformation. This block-diagonalizes the system into eight irreducible representations for each value of angular momentum. For most cases, we study systems on graphs that do not exhibit rotational symmetry for the sake of generality. 

\begin{figure}
    \includegraphics[width=0.45\textwidth]{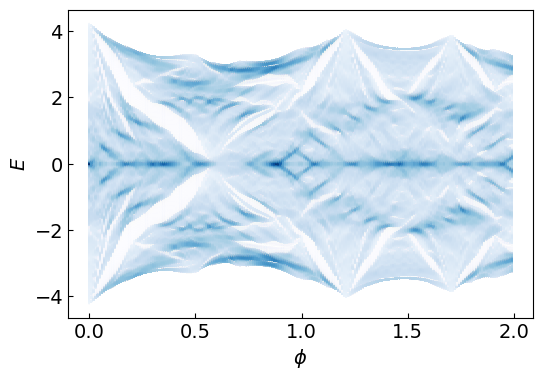}
    \captionsetup{justification=centerlast} 
    \caption{ Spectrum of the Hofstadter model a circular cutout of the AB-tiling with radius $34$ centered at the point with coordinates $(17,16)$ to avoid rotational symmetry. This system consists of $8772$ sites. The spectrum is calculated for $\phi \in (0,2)$ where darker colors symbolize larger density of states. Due to chiral symmetry the energy spectrum is mirror symmetric around $E=0$.}
    \label{fig:energy_spectrum}
\end{figure}

Independent of the choice of boundary, the systems exhibit chiral symmetry, which emerges due to the bipartite nature of the graph. This symmetry is represented by a Hermitian and unitary operator $\Gamma$ satisfying $\Gamma H \Gamma^\dagger = -H$. This symmetry ensures that the spectrum of the Hamiltonian is mirror symmetric around $E=0$, see Fig.\ \ref{fig:energy_spectrum}. For a discussion of bipartite graphs and a derivation of the chiral symmetry operator, see appendix \ref{appendix:sym}.

\subsection{Adiabatic Charge Pump and Imperfect Topology}

When describing the topology of the Hofstadter model, one usually uses the Chern number. This is commonly calculated as the integral of the Berry curvature over the Brillouin zone for each band in the bulk Hamiltonian \cite{TIintro}. However, this is not possible for our model as the AB-tiling does not have translational invariance, so momentum is not a good quantum number. This necessitates the use of other methods for calculating the Chern number that are based on real space rather than momentum space.

\subsubsection{Adiabatic Charge Pump}

One method for measuring topology is to use an adiabatic charge pump as done in Refs.\ \cite{Tong, pump}. In this method, we introduce a flux tube localized at a single point in the $xy$-plane which we pick to be the origin. We then vary the flux $\varphi_c$ given in units of the flux quantum $\phi_0 = 2 \pi$ from $\varphi_c = 0$ to $\varphi_c = 1$. Let us therefore write the flux as $\varphi_c(t) = t/T$, with $T$ being a timescale chosen such that we stay in the adiabatic regime. From Stoke's theorem this yields a magnetic potential which in cylindrical coordinates is given by $\textbf{A} = (\varphi_c / r)\hat{\phi}$. This enters the Hamiltonian as an additional Peierls phase, which can be calculated using independence of integration path for conservative vector fields. Let $z_j = x_j + iy_j$ denote the position of site $\textbf{r}_j$ as a complex number. Then the Peierls phase can be written as
\begin{equation}
    \theta_{ij} = \varphi_c \arg \left( \frac{z_j}{z_i} \right),
\end{equation}
where $\arg(z) \in (-\pi, \pi]$. Whenever $\varphi_c = n$ for $n \in \mathbb{Z}$, the Peierls phase can be gauged away using the gauge transformation: $\ket{\textbf{r}_i}\rightarrow \ket{\textbf{r}_i} \exp(i \ n \arg(z_i))$. Therefore, the spectrum of our Hamiltonian is the same whenever the flux is an integer multiple of the flux quantum. \smallskip

As we adiabatically change the flux from zero to one unit of the flux quantum, we start and end with the same spectrum. However, during this process, occupied states below the Fermi energy may be pumped across the Fermi energy to a previously unoccupied state or vice versa. Among the states with real-space localization within a radius $r$ from the flux tube, let $N_+(r)$ denote the number of states pumped from below to above the Fermi energy during a cycle and let $N_-(r)$ be the number of states pumped from above to below. Then the Chern number at a radius $r$ from the flux tube can be expressed as
\begin{equation}\label{eq:pump}
    C(r) = N_-(r)-N_+(r).
\end{equation}
For this equation to work, all states pumped past the Fermi energy that are not counted in the above expression should be localized outside the radius $r$. This ensures that they do not overlap with the counted states.

\subsubsection{Imperfect topology}

Consider the case of a Chern insulator on a two-dimensional crystal with periodic boundary conditions along the $y$-axis and open boundary conditions along the $x$-axis. If the boundaries on the $x$-axis are close together, the band structure exhibits an avoided crossing in the bulk gap. As we approach the thermodynamic limit, this avoided crossing turns into a crossing protected by the real-space distance of the edge states. This makes it easy to define when the system is topological by analyzing it in the thermodynamic limit.

\begin{figure}
    \includegraphics[width=0.45\textwidth]{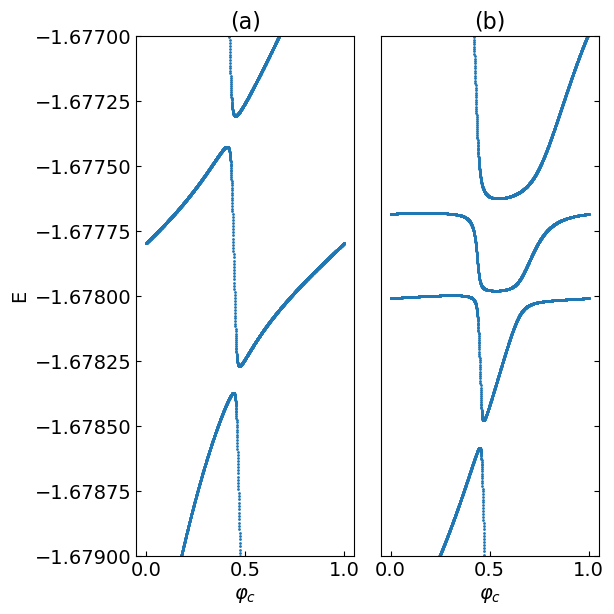}
    \captionsetup{justification=centerlast} 
    \caption{Spectral flow of the model with $\phi=0.7375$ chosen in accordance with later results. The graph is a circular cut-out whose origin has been shifted by the vector $(x,y)=(17,16)$ to avoid rotational symmetry. (a) Spectral flow on a graph of radius $24$ corresponding to $4364$ sites. (b) Spectral flow on a graph of radius $34$ corresponding to $8772$ sites. The states shown in graph (b) are among those later identified as BLT states. Both graphs show an avoided crossing but the size of the gap is the same for both systems.}
    \label{fig:imperfect}
\end{figure}

This is not applicable for the case of BLT states, where the topological current is located inside the bulk of the graph. Because of this, there is no reason to expect that they exhibit perfect crossings in the limit of infinite system sizes, since they can be localized at many different radii independent of system size, see Fig.\ \ref{fig:imperfect}. This means there is no limit in which the system shows perfect topology. However, this still holds experimental relevancy, as we still get close to integer currents that exhibit high levels of robustness. Furthermore, topological edge states are only perfectly topological in the limit of infinite system sizes, which is not physically realizable. We therefore need some approximate definition of what we consider to qualify as a BLT state. To this end, we introduce a marker that can identify the localization of the topological current and is built on physically measurable quantities.  

\subsection{The Crosshair Marker}

The main method we use to measure topology is a local measure of the Chern number called the crosshair marker, which is calculated for each site in the graph, derived in Ref.\ \cite{crosshair}. The method is based on applying an electric field that is only non-zero on the line $y=R_y$ given by
\begin{equation}
    \textbf{E}(\textbf{r}) = V_0 \delta(y-R_y) \hat{y}.
\end{equation}
The current is then measured across a line given by $x=R_x$, such that the electric field and the line of measurement form a crosshair centered at $\textbf{R} = (R_x,R_y)$. Around the location of the crosshair we get contributions from the Hall current which for topological systems sum to the Chern number. Further from the crosshair, we get current contributions from where the system turns conducting. These currents are exactly the topological currents associated with the topological invariant. This is in accordance with the bulk-boundary correspondence, but for our case, we find topological currents localized away from the edge. The expression of the crosshair marker for a site $\textbf{r}$ is given by
\begin{equation}\label{eq: crosshair}
    C_{\textbf{R}}(\textbf{r}, E) = 4\pi \Im \Tr_\textbf{r}{\left(P(E)\vartheta_{R_x}P(E)\vartheta_{R_y}P(E)\right)},
\end{equation}
where $\Tr_{\textbf{r}}{(...)}=\Tr(\ket{\textbf{r}}\bra{\textbf{r}}...)=\bra{\textbf{r}}...\ket{\textbf{r}}$. \newline $\vartheta_{R_x} = \sum_i \theta(x_i-R_x)\ket{\textbf{r}_i}\bra{\textbf{r}_i}$ denotes the projection operator onto the right side of the graph, with $\vartheta_{R_y}$ defined analogously. Finally, $P(E) = \sum_{E_i\leq E} \ket{\psi_i}\bra{\psi_i}$ denotes the projector onto the occupied eigenstates for a Fermi energy $E_f = E$. Throughout this article, whenever we refer to some state, $i$, we refer to the many-body ground state where all states of energy less than or equal to $E_i$ are occupied described by the projection operator $P(E)$. For the full derivation of the crosshair marker, see Ref.\ \cite{crosshair}.

This marker proves a powerful tool, as it allows us to determine exactly which sites on the graph support currents for a given Fermi energy, allowing us to distinguish between edge currents and bulk localized currents. Furthermore, the marker distinguishes itself from other markers such as the Bott index by being based on physical measurable quantities, such that the result in all regimes has intuitive meaning.

A powerful consequence of chiral symmetry is that if we have a topological current at a Fermi energy $E_f$, then there must be a corresponding topological current at energy $-E_f$ localized in the same area of the graph but with the opposite Chern number. Mathematically this can be expressed using the crosshair marker as
\begin{equation}
    C_{\textbf{R}}(\textbf{r}, -E) = -C_{\textbf{R}}(\textbf{r}, E).
\end{equation}
From this result, we can simply count all states that correspond to positive Chern numbers. Then, by symmetry, there must be an equal amount of states corresponding to negative Chern numbers. For a derivation of this formula, see appendix \ref{appendix:cross}. 

\section{Procedure and Measures}\label{sec:alg}

Now that we have gone through the most important aspects of the theory, we can define the procedure used to identify topological states in our system. The systems we analyze consist of the Hofstadter model on a circular cutout of the graph based on a given origin. Therefore, any model we consider is uniquely defined by the radius, flux, and choice of origin, $(R_0, \phi, x_0, y_0)$. Here, $\phi$ refers to the flux in units of the flux quantum. $(x_0, y_0)$ is the choice of origin measured from the rotational symmetry axis with respect to the graph orientation shown in Fig.\ \ref{fig:AB}. For any specific cut out we work in a new coordinate system where the origin $(0,0)$ coincides with $(x_0, y_0)$ in the old coordinate system. In the following sections, all considered models are of this type. 

\subsection{Identification Procedure}

The goal of the identification procedure is to create a numerical quantitative technique to identify the topological states. The procedure should be based on measurable quantities which are also numerically inexpensive to compute. To this end, we apply the crosshair marker, while the adiabatic charge pump serves as a complementary tool for qualitative comparison. Since the topology in BLT states is not perfect as discussed above, we will introduce parameters which quantify how close to topological a state needs to be to be flagged as topological by the numerical procedure. By making these parameters stricter one will generally get fewer states of higher quality. We will also introduce a measure to define how far from the edge a state should be to be considered a BLT state rather than an ELT state.

Assume for all future calculations that the crosshair is located at the origin of the graph, $\textbf{R}=(0,0)$. As such, for convenience, we omit $\textbf{R}$ from the notation in the remainder of the article. Let us now define the cumulated crosshair marker for a given Fermi energy by
\begin{equation}
    \Lambda(r, E) = \sum_{\norm{\textbf{r}}\leq r}C(\textbf{r}, E).
\end{equation}

\begin{figure}
    \includegraphics[width=0.45\textwidth]{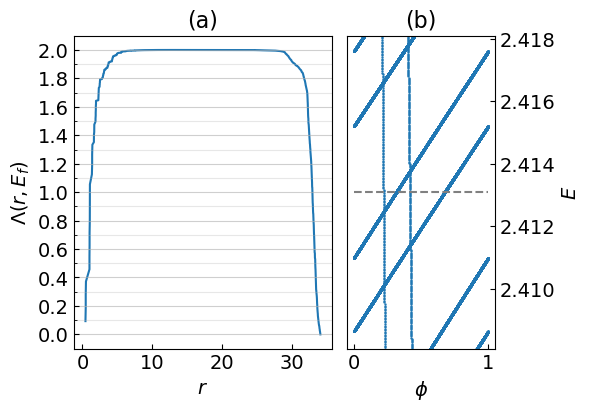}
    \captionsetup{justification=centerlast} 
    \caption{ Visualization of topological measures for a square lattice of $8586$ sites cut out in a circle of radius $34$ and flux $\phi=0.69$ at a Fermi energy $E_f=2.413$. (a) The cumulated crosshair marker $\Lambda(r, E_f)$ as a function of radius. The plateau of the graph indicates a Chern number of two. (b) Spectral flow of the adiabatic charge pump showing two states transported past the Fermi energy highlighted as a dashed line in accordance with (a).}
    \label{fig:square}
\end{figure}

Note that summing over all sites in the system, $r=R_0$, yields $\Lambda(R_0, E)=0$. This is exactly the marker defined by Kitaev \cite{Kitaev}, which turns out to be zero for finite systems with open boundary conditions. If a system is topological, we expect $\Lambda(r, E)$ will achieve a maximal value equal to a non-zero integer. This leads us to define the quantity
\begin{equation}
    Q(E) = \max_r \  \Lambda(r, E)
\end{equation}
For a square lattice, it can be seen how $Q(E)$ corresponds to the Chern number for the Fermi energy $E$, as $Q(E)$ exactly equals the number of pumped states for the adiabatic charge pump, see Fig.\ \ref{fig:square}. 

One only needs to check if $Q(E)$ is an integer for each eigenvalue $E_i$ of the Hamiltonian, as each eigenvalue corresponds to a different number of occupied states. As we work numerically and with systems that do not exhibit perfect topology, we cannot expect states to have $Q(E_i)$ exactly equal to an integer. Therefore, we need to check whether $Q(E_i)$ falls within some small interval centered at an integer. From this we can define the set of indices $i$ such that the eigenvalue $E_i$ of $H$ corresponds to a Fermi energy, where the system can host topological states as
\begin{equation}\label{eq:top_index}
    S = \left\{ i \mid Q(E_i) \in (n-\epsilon, n+\epsilon) \mbox{  for some  } n\in \mathbb{Z}/\{0\} \right\}.
\end{equation}

Throughout our calculations, we let $\epsilon = 0.01$. One should be careful when using this method, as there can be non-topological states, which by coincidence have $Q(E_i)$ close enough to an integer to fall within $S$. Being an element in $S$ is therefore a necessary but not sufficient requirement for a Fermi energy to correspond to a topological state.
To address this, we may introduce disorder in the system and record the states from $S$ where the system can host robust currents. As done in Ref.\  \cite{Saswat}, we introduce a disorder term to our Hamiltonian consisting of an additional on-site potential for each site in the graph. The total Hamiltonian, $H_{\text{dis}}$, is then given by
\begin{equation}
    H_{\text{dis}} = H + \sum_{j} \alpha_j(W) \ket{\textbf{r}_j}\bra{\textbf{r}_j},
\end{equation}
where $\alpha_j(W)$ is a number drawn from a uniform distribution on $[-W/2, W/2]$, with $W$ denoting the magnitude of the disorder. Let $Q_{\text{dis}}(E_i)$ denote the maximal value of the cumulated crosshair marker for $H_{\text{dis}}$ analogous to the definition of $Q(E_i)$, but with $E_i$ still referring to eigenvalues of the original Hamiltonian, $H$. We then want to pick some $W$ where we can distinguish between topological and non-topological states using $Q_{\text{dis}}(E_i)$. If $W$ is too small, it can be hard to distinguish between these cases, and if $W$ is too large, then too much information is lost. 

\begin{figure}
    \includegraphics[width=0.45\textwidth]{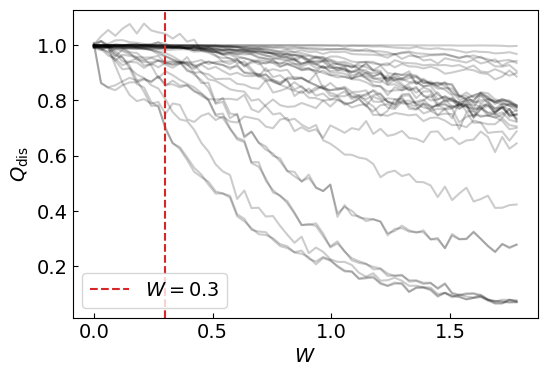}
    \captionsetup{justification=centerlast} 
    \caption{Maximal value of the cumulated crosshair marker $Q_{\text{dis}}(E_i)$ as a function of the strength of disorder $W$ for a sample of 35 states. The calculations are done for the system $(R_0, \phi, x_0, y_0) = (34, 0.69, 17, 16)$. At low $W$ the non-robust states are too close to the robust ones to distinguish accurately. At high $W$ the states bundle together according to energy due to the disorder erasing finer details. $W=0.3$ marks the chosen disorder between these limiting cases.}
    \label{fig:disorder}
\end{figure}

To choose an appropriate disorder magnitude $W$ we compute $Q_{\text{dis}}(E)$ for a range of $W$ in some sample system. Here we picked the system $(R_0, \phi, x_0, y_0) = (34, 0.69, 17, 16)$, which corresponds to a graph of 8772 sites. The flux was chosen according to the choice from Ref.\ \cite{A2}, and the offset of the origin is to avoid rotational symmetry. For a random sample of states in $S$, $Q_{\text{dis}}(E_i)$ is calculated for each disorder $50$ times, and the average is calculated. The result is seen in Fig.\ \ref{fig:disorder}, which leads us to choose $W=0.3$ for future robustness calculations. As seen in the figure, all the sampled states correspond to a Chern number of one. This is because this system only exhibits a maximal Chern number of one, and chiral symmetry allows us to only consider positive Chern numbers. As seen later in the article, systems exhibiting larger Chern numbers are scarce and play a minimal role in further computations. 

Now that we have determined an appropriate disorder for checking robustness, we can compute $Q_{\text{dis}}(E_i)$ for all $i \in S$ at $W=0.3$. This is done $200$ times and the average $\mu_{i}$ and standard deviation $\sigma_{i}$ is found for each energy. Let $n_i \in \mathbb{Z}/\{0\}$ denote the integer such that $Q_{\text{dis}}(E_i) \in (n_i-\epsilon, n_i+\epsilon)$ for $i \in S$. Then we may define a robustness measure, $\Omega_{i}^Q$, by
\begin{equation}
    \Omega_{i}^Q = \abs{n_i-\mu_{i}}+\sigma_{i}.
\end{equation}

The closer the marker is to zero, the greater the robustness of the current. The standard deviation is added because some currents will, like edge currents, have a mean close to $n_i$. They are nevertheless distinguishable from edge currents due to their high standard deviation. We want our marker to take these factors into account as it implies a greater instability towards disorder in our system. With this we can now determine which states are robust and which are not.  

Finally, we need a way to distinguish between
BLT and ELT states. To do this, we look at how much support the probability density of a state has by the edge, which can vary smoothly from state to state. This necessitates introducing a threshold to create a binary classification. Therefore, we construct a radial measure $\Omega_{i}^R$ for some $i\in S$. 
\begin{align}
    \Omega_{i}^R & = \max A_i/ \max \left( \bigcup_{\{j \in S \ : \ n_j=n_i\}} A_j \right), \\
    A_i & = \{ r \mid \Lambda(r, E_i) > n_i-0.05 \}.
\end{align}
This measures at what radius $\Lambda(r, E_i)$ falls below a certain threshold, which is where the topological current is localized. The marker is normalized with respect to the maximal radius for all the indices $i \in S$ that correspond to the same Chern number. With these two markers, we can finally define the subsets, $S_B$ and $S_E$, of $S$ corresponding to the topological bulk states and the edge states, respectively
\begin{align}
    S_B & = \left\{ i\in S \mid  \Omega_{i}^Q<0.05, \quad \Omega_{i}^R < 0.97 \right\} \\
    S_E & = \left\{ i\in S \mid  \Omega_{i}^Q<0.05, \quad \Omega_{i}^R > 0.97 \right\}
\end{align}
The thresholds used in these definitions were qualitatively determined using data visualizations.

\subsection{Counting Edge States}

From the procedure, we can count the number of topological edge states as $\abs{S_E}$. This number has to be close to the number of edge states in the system. We therefore define a method for identifying the edge states in order to check if the numbers match. Let $\mathcal{R}$ denote the operator that measures the radial localization of a site $\textbf{r}$ with respect to the origin
\begin{equation}
    \mathcal{R} \ket{\textbf{r}} = \norm{\textbf{r}} \ket{\textbf{r}}.
\end{equation}
From this we can define a radial measure on the $i$th eigenstate $\ket{\psi_i}$ as
\begin{equation}
    M_{\mathcal{R}}^{i} = \bra{\psi_i} \mathcal{R} \ket{\psi_i}/\max_j (\bra{\psi_j} \mathcal{R} \ket{\psi_j})
\end{equation}

\begin{figure}
    \includegraphics[width=0.45\textwidth]{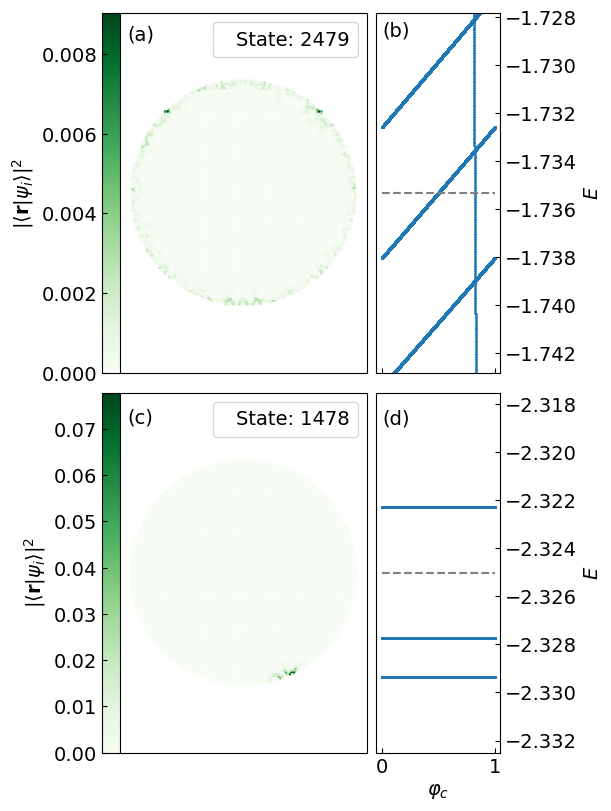}
    \captionsetup{justification=centerlast} 
    \caption{The spectral flow of two kinds of edge states in the system $(R_0, \phi, x_0, y_0) = (34, 0.69, 17, 16)$. (a) Density plot of an edge state distributed close to evenly around the perimeter of the graph. (b) The spectral flow of the edge state in (a) showing a Chern number of one. (c) Density plot of an edge state localized on a small area of the perimeter. (d) The spectral flow of the edge state in (c) showing no transport across the Fermi energy corresponding to a Chern number of zero}
    \label{fig:edge}
\end{figure}

This measure can identify whenever a state is localized on the edge, but it cannot distinguish between the two kinds of edge states; those that are localized all around the edge and those that are localized only on a certain area of the edge. The latter kind of edge state does not contribute to topological edge states as it does not enclose the crosshair. This is also seen by application of the adiabatic charge pump, as they do not experience any charge transport, see Fig.\ \ref{fig:edge}. 

To distinguish between these states, we can divide the graph into four quadrants and then demand that the edge state is present in each quadrant. As such, define the projector onto the $m$th quadrant $\vartheta_{m}$ as
\begin{align}
    \vartheta_{m} & = \sum_i \theta(s_x^m x_i)\theta(s_y^m y_i)\ket{\textbf{r}_i}\bra{\textbf{r}_i}, \\
    s_x^m & = \mbox{sgn}\left[\cos\left(m\frac{\pi}{2}-\frac{\pi}{4}\right)\right],\\
    s_y^m & = \mbox{sgn}\left[\sin\left(m\frac{\pi}{2}-\frac{\pi}{4}\right)\right],
\end{align}
where $s_x^m$ and $s_y^m$ yield the sign of the $x$ and $y$ coordinates as a function of the quadrant number. We can now define the set of eigenstate indices, $i$, of the Hamiltonian, where we have an edge state contributing to topological edge states as
\begin{align}
    S_\psi = \{i \mid & M_{\mathcal{R}}^{i}>0.97, \notag \\ 
    & \quad \bra{\psi_i} \vartheta_{m} \ket{\psi_i}>0.2 
 \ \ \forall m\in\{1,2,3,4\}\}.
\end{align}

The threshold for $M_{\mathcal{R}}^{i}$ was chosen in accordance with the threshold for $\Omega_{i}^R$ such that the two sets are comparable. For the system, $(R_0, \phi, x_0, y_0)=(34, 0.69, 17, 16)$, the number of edge states has been calculated to be $\abs{S_\psi} = 68$ while the number of topological edge states are $\abs{S_E} = 62$ which are comparable in size. 
The difference between these numbers is due to the fact that each measure describes different things. One describes the localization of the ELT state while the other is the localization of the induced current at the corresponding Fermi energy.

\section{Data Collection}\label{sec:data}

We now use the procedure to identify topological states in different systems. 

\subsection{Variable Flux}

\begin{figure}
    \includegraphics[width=0.45\textwidth]{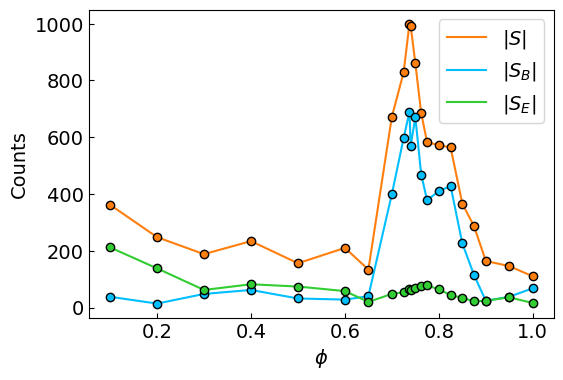}
    \captionsetup{justification=centerlast}
    \caption{ The number of initial states $\abs{S}$, BLT states $\abs{S_B}$, and topological edge states $\abs{S_E}$ as a function of the flux for the systems $(R_0, \phi, x_0, y_0)=(34, \phi, 17, 16)$. This shows the only instances of significant numbers of BLT states occur at $\phi \in (0.65, 0.9)$}
    \label{fig:flux}
\end{figure}

We start by computing the number of topological states as a function of the flux. Let us therefore analyze systems of the form $(R_0, \phi, x_0, y_0)=(34, \phi, 17, 16)$ where we vary $\phi$ from zero to one in units of the flux quantum. Using the procedure from the previous section, we find that the number of topological states spikes in the area $\phi \in (0.65, 0.9)$ while having few topological states outside this interval, see Fig.\ \ref{fig:flux}. Note that the spectrum of the Hamiltonian is not periodic in $\phi$ so these measurements do not conclude anything beyond a flux of $\phi_0$. 

Among the data points, the highest number of bulk localized topological states was found at $\phi = 0.7375$, which is the flux we use for all computations below. The only Chern numbers observed for these systems were at most one except for low fluxes $\phi = 0.1, 0.2$. At $\phi = 0.1$ Chern numbers up to three were observed, but the only topological states found with Chern numbers greater than one were edge states. For $\phi = 0.2$ Chern numbers up to two were found, but no BLT states greater than one were found. 

In Fig.\ \ref{fig:flux} the number of edge states at $\phi = 0.1, 0.2$ is greater than those found at larger fluxes. This is because larger Chern numbers correspond to edge currents that penetrate deeper into the material. As such, more eigenstates are involved in the currents leading to a larger total number of edge states.

\subsection{Variable Radius}

\begin{figure}
    \includegraphics[width=0.45\textwidth]{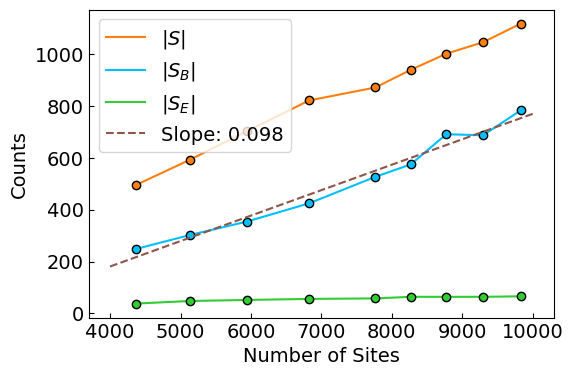}
    \captionsetup{justification=centerlast}
    \caption{ The number of initial states $\abs{S}$, BLT states $\abs{S_B}$, and topological edge states $\abs{S_E}$ as a function of the number of sites for the systems $(R_0, \phi, x_0, y_0)=(R_0, 0.7375, 17, 16)$. The number of BLT states follow a linear tendency in the number of sites, $N$, given by $N_\text{BLT} \approx 0.098N-210$.}
    \label{fig:radius}
\end{figure}

We now look at how the number of BLT states is impacted by the number of sites in our graph. As such, consider systems of the kind $(R_0, \phi, x_0, y_0)=(R_0, 0.7375, 17, 16)$ from which we can plot the number of BLT states as a function of radius as seen in Fig.\ \ref{fig:radius}. 

All topological states found in this system correspond to Chern numbers of at most one. The number of BLT states follows a linear tendency and can, through linear regression in the number of sites $N$ be approximated by $N_{BLT} \approx 0.098N-210$. While the number of edge states is usually proportional to the radius of the system, these topological states show proportionality to the system size or area of the system. This further reinforces the fact that being localized in the bulk is an integral part of their nature and they scale according to two spatial dimensions as opposed to edge states confined to one dimension.

\section{Characteristics of BLT states}\label{sec:BLT}

Until now, we have been discussing how to identify the presence of BLT states. A natural extension of this analysis is to examine which properties these states exhibit. 

\subsection{Density of States}

\begin{figure}
    \includegraphics[width=0.45\textwidth]{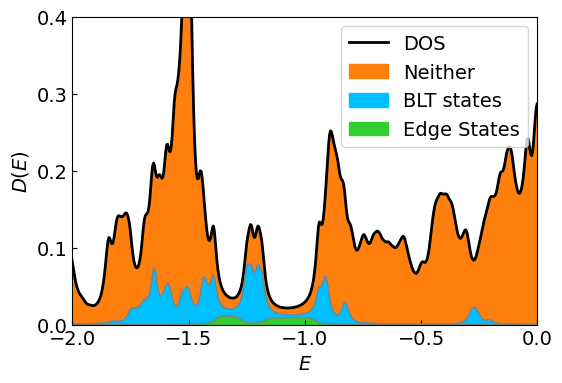}
    \captionsetup{justification=centerlast}
    \caption{  Density of states calculated for the system $(R_0, \phi, x_0, y_0)=(34, 0.7375, 0.1, 0.3)$ using $\epsilon=0.016$ in Eq.\ (\ref{eq:density}). Each colored area corresponds to how large the contribution from a given kind of state is to the total value of $D(E)$. Edge states are localized in areas of small values of $D(E)$ while BLT states can appear at both high and low values of $D(E)$.}
    \label{fig:density}
\end{figure}

Edge states are found in areas of the energy spectrum with low density of states. As such, one might imagine analogous properties to hold for BLT states. To calculate the normalized density of states for a given energy, $E$, we use the approximation
\begin{align}\label{eq:density}
    D(E) & = \frac{1}{N} \sum_i^N \delta(E-E_i) \notag \\
    & \approx \frac{1}{N \pi } \sum_i^N \frac{\varepsilon}{(E-E_i)^2 + \varepsilon^2},
\end{align}
where $N$ denotes the number of eigenstates of the Hamiltonian $H$ and $\varepsilon$ is some small number corresponding to the width of the peaks located at each energy $E_i$. We can then calculate the density of states for an interval of energies for the system $(R_0, \phi, x_0, y_0)=(34, 0.7375, 0.1, 0.3)$. To this end, let $\varepsilon \approx 1.6\cross 10^{-2}$ which is $50$ times the median of the energy gaps in the spectrum. This value is chosen such that the density of states at a given point picks up close lying states but not so big as to erase the finer details. We can then calculate the contribution from the BLT states or the edge states by restricting the sum in Eq.\ (\ref{eq:density}) to $i \in S_B$ or $i \in S_E$, the result of which is seen in Fig.\ \ref{fig:density}. 

From Fig.\ \ref{fig:density} one sees that unlike edge states, BLT states are not limited to a low density of states. Furthermore, there are states in the set $S$ that are neither BLT nor edge, yet reside in regions of low density of states. As such, other factors may play a role in preserving the robustness of the BLT states.  

\subsection{Geometric Support and Localization}

\begin{figure*}[ht]
\captionsetup[subfloat]{position=top}
\centering
  \subfloat[]{  
    \includegraphics[width=0.45\textwidth]{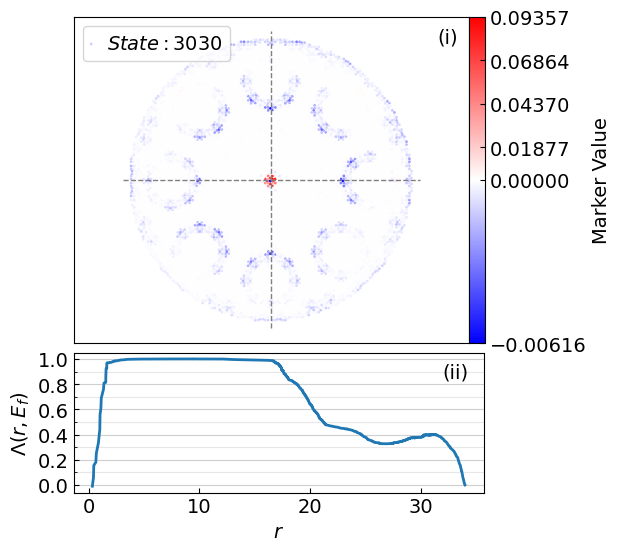}%
    \label{fig:current1}%
    }%
  \subfloat[]{%
    \includegraphics[width=0.45\textwidth]{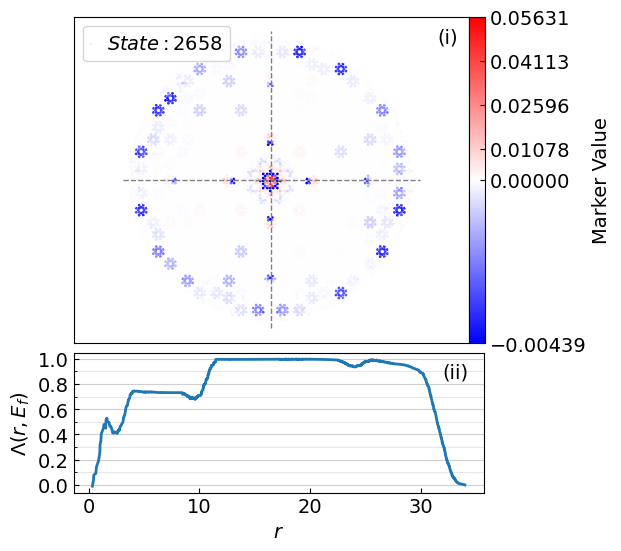}%
    \label{fig:current2}%
  }
%\captionsetup{justification=centerlast}
\caption{ Visualization of topological currents using the crosshair marker corresponding to the system $(R_0, \phi, x_0, y_0)=(34, 0.7375, 0.1, 0.3)$ for the states: (a) $3030\in S_B$. (b) $2658\in S_B$. (i) Density plot of the crosshair marker with the dashed lines indicating the position of the crosshair. Contributions from Hall conductance is seen at the location of the crosshair, and away from the crosshair contributions are present where the system turns conducting. This represents the localization of the topological current. (ii) The cumulated crosshair marker $\Lambda(r, E_f)$ as a function of radius. The plateau for each graph indicates a Chern number of one. At (a) the topological currents traces circular patterns corresponding to sites of coordination number seven. At (b) the contributions to the topological current are localized close to sites of coordination number eight.}
\label{fig:fig}
\end{figure*}

Since it is the aperiodic nature of the graph that induces the existence of BLT states, one may ask if the geometry of the graph can be correlated to the properties of the BLT states such as shape and localization. One property of the AB-tiling, that distinguishes itself from the usual square lattice, is the number of connections (coordination number) for a given site $\textbf{r}_i$. For a square lattice, this number is always four, but for the AB-tiling it can be up to eight. As such, we can check if some topological currents have support which follows the pattern traced by sites belonging to specific coordination numbers. In order to better illustrate the shapes of the BLT states we look at a system slightly offset from the point of rotational symmetry but enough to change the graph, $(R_0, \phi, x_0, y_0)=(34, 0.7375, 0.1, 0.3)$. For this system, we can find many BLT states that have support on sites of coordination number seven resulting in circular patterns as seen in Fig.\ \ref{fig:current1}. 
\begin{figure}
    \includegraphics[width=0.45\textwidth]{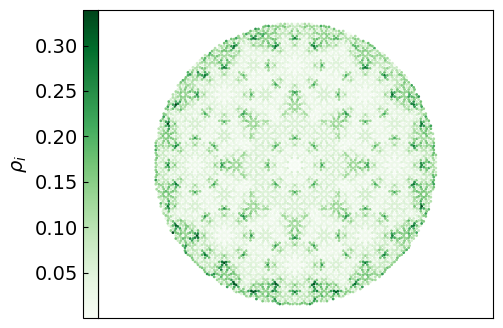}
    \captionsetup{justification=centerlast}
    \caption{The sum of the probability density for all BLT states in $S_B$ for the system $(R_0, \phi, x_0, y_0)=(34, 0.7375, 0.1, 0.3)$ calculated for each site of the graph. This shows localization in specific areas of the graph following the patterns traced by sites of coordination number seven or eight.}
    \label{fig:BLT_density}
\end{figure}
The number of states with this kind of support depends on what graph cut-out and flux we are using, but they are nevertheless present in many systems. 

There are also numerous currents that do not fit this pattern, one of the common kinds being shown in Fig.\ \ref{fig:current2}. This pattern consists of smaller circles, each of which is centered at a point of coordination number eight. These are instances of currents that follow distinguishable patterns, but there are also many that do not. However, in the systems we have analyzed almost all the BLT states that appear at higher density of states follow one of these patterns. As such, the robustness of BLT states at higher density of states could be attributed to their support on certain patterns. 

To illustrate the localization of all the BLT states at once, we can create a measure that sums the probability density for each eigenstate, $\ket{\psi_j}$, of $H$ with $j\in S_B$ with respect to a given site in the graph. For the site $\textbf{r}_i$ the measure can be written as
\begin{equation}
    \rho_i = \sum_{j\in S_B} \abs{\braket{\textbf{r}_i}{\psi_j}}^2
\end{equation}

Using this, we can see if there is a general bias towards BLT states having support on certain areas of the graph, the result of which is seen in Fig.\ \ref{fig:BLT_density}. This shows that, in general, the BLT states follow certain patterns in the graph, which could induce greater robustness. Specifically, we see circular patterns and points in accordance with the graph sites of coordination number seven and eight. This is the case for this system, but other systems can exhibit more disordered shapes making potential patterns harder to identify.

\subsection{Robustness and Variation}

\begin{figure*}[ht]
  \centering
  \captionsetup[subfloat]{position=top}

  \subfloat[]{  
    \includegraphics[width=0.45\textwidth]{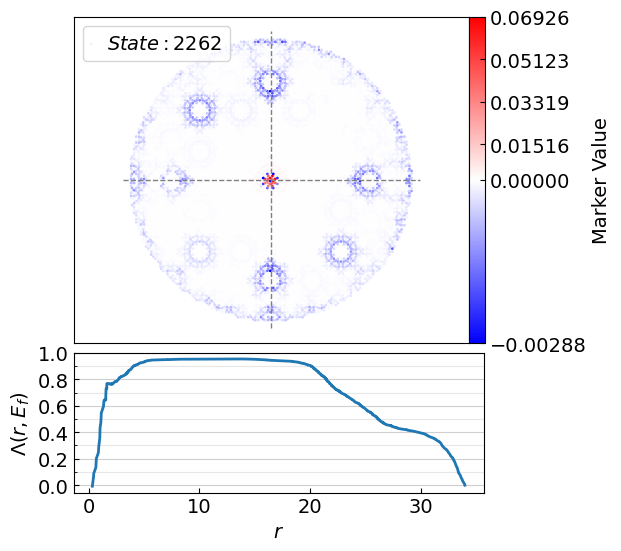}%
    \label{fig:nontop1}%
    }%
  \subfloat[]{%
    \includegraphics[width=0.45\textwidth]{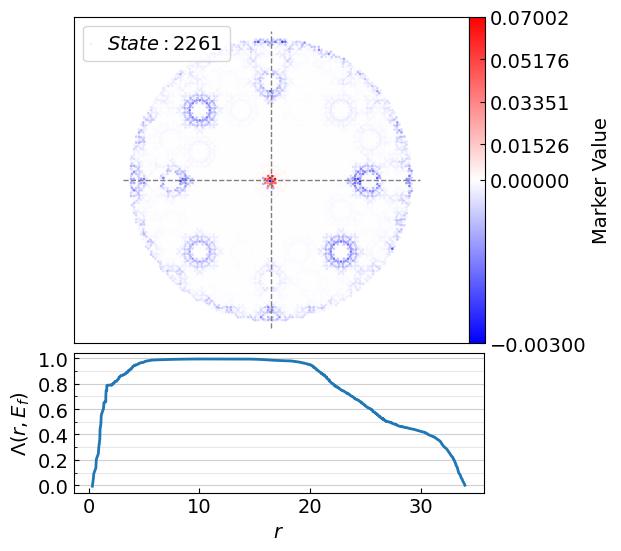}%
    \label{fig:top1}%
  }
  \vfill
  \subfloat[]{  
    \includegraphics[width=0.45\textwidth]{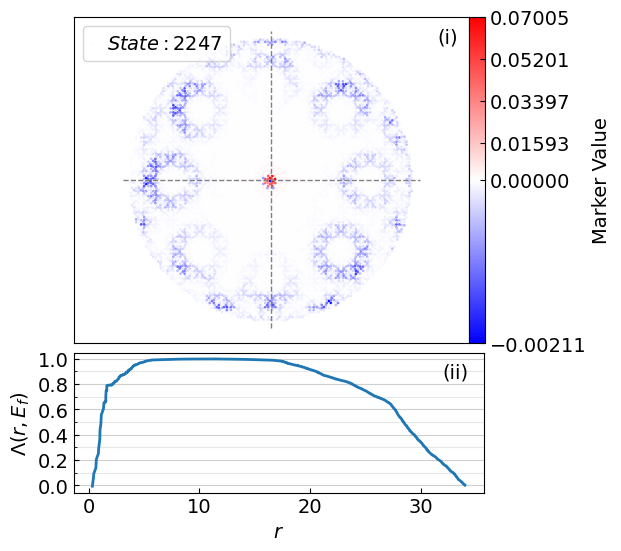}%
    \label{fig:BLT1}%
    }%
  \subfloat[]{%
    \includegraphics[width=0.45\textwidth]{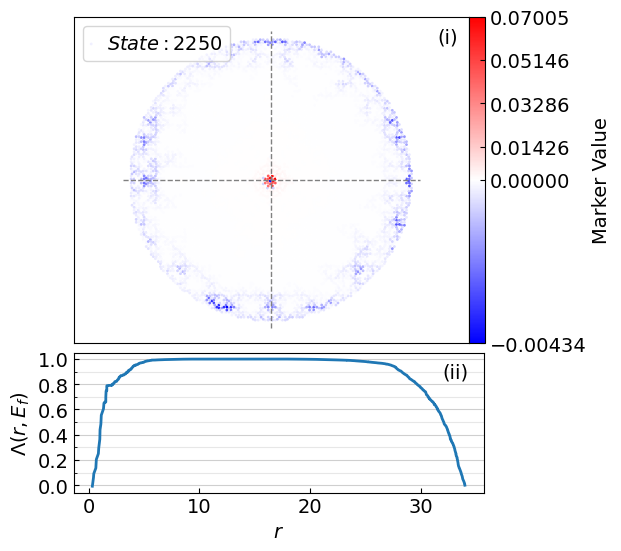}%
    \label{fig:BLT2}%
  }
\caption{ Visualization of topological currents using the crosshair marker corresponding to the system $(R_0, \phi, x_0, y_0)=(34, 0.7375, 0.1, 0.3)$. (i) Density plot of the crosshair marker with the dashed lines indicating the position of the crosshair. (ii) The cumulated crosshair marker $\Lambda(r, E_f)$ as a function of radius. The figure in (a) represents a state that was not sufficiently topological to be picked out by the procedure as a BLT state. This is due to the maximal value of the cumulated crosshair marker lying outside the interval described in eq.\  (\ref{eq:top_index}). Nevertheless, it closely resembles the figure in (b) which corresponds to a BLT state. (c) and (d) represents two BLT states within the same group of states. This illustrates how the localization of the topological current may change throughout the group of BLT states without changing the Chern number.}
\label{fig:border}
\end{figure*}

An interesting point to study is whether the BLT states through disorder can transition to non-topological states. Edge states are protected as they appear in large groups consisting of consecutive edge states distributed over large spans of energy. As such, it would take a big energy leap to move from an edge state inside the group to a non-edge state. Previously we have shown that BLT states do not need to appear at low density of states, but it is worth checking if they appear in groups, and how large the energy span of these groups are. 

To this end, we consider the system $(R_0, \phi, x_0, y_0)=(34, 0.7375, 0.1, 0.3)$. All of the topological edge states in $S_E$ with Chern number one belong to one of two groups which can also be seen in Fig.\ \ref{fig:density}. We define the energy span $\Delta E$ of a group of states as the energy difference between the bordering states that do not belong to the group. The energy spans of the groups of topological edge states are $\Delta E = 0.127$ and $\Delta E = 0.227$ and consist of thirteen and nineteen states respectively. 

For the case of BLT states, $S_B$, they can appear both in groups and individually. If we define a group of states to consist of consecutive BLT states with $\Delta E > 0.01$, then $380$ of the total $592$ BLT states belong to one of such groups. Among these, the largest energy span is $\Delta E = 0.0583$ consisting of fifteen states which is approximately half the energy span of the smallest group of edge states. 

We can then check if the non-topological states bordering these groups resemble the BLT states identified by the procedure. If they do, the states by the edges of the group are less likely to be robust as they can be more easily perturbed into non-topological states. This also implies that smaller groups would be less likely to be topological when subjected to different types of disorders. We therefore apply the crosshair marker to compare how much the induced currents resemble each other. For almost all cases, they resemble each other to a large degree while being close in energy, an example of such seen in Figs.\ \ref{fig:nontop1}, \ref{fig:top1}. The close resemblance implies that BLT states that lie completely isolated from other BLT states could more easily transition to a state the procedure considers to be non-topological. 
Within a group of BLT states, the shape of the topological current may also change. This is illustrated in Figs.\ \ref{fig:BLT1}, \ref{fig:BLT2}, depicting the topological current from two states of the same group that are localized in different areas of the graph. 

\section{Discrepancies and Possible Amendments}\label{sec:Discuss}

\begin{figure}
    \includegraphics[width=0.45\textwidth]{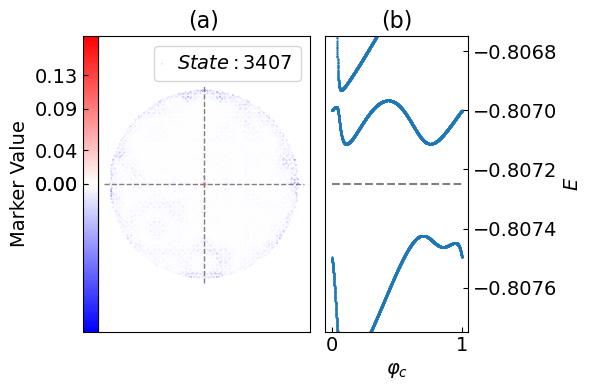}
    \captionsetup{justification=centerlast} 
    \caption{ Visualisation of topological techniques for the state $3407 \in S_B$ in the system $(R_0, \phi, x_0, y_0)=(34, 0.7375, 17, 16)$. (a) Density plot of the crosshair marker, with the dashed line indicating the position of the crosshair. The topological currents is shown in blue away from the crosshair where the system turns conducting. The current is smeared not adhering to any obvious pattern. (b) Visualisation of the adiabatic charge pump at Fermi energy just above the energy of state $3407$. There is no charge flow across the Fermi level which could contradict the state being identified as topological. Whether the state can jump the avoided crossing, thus avoiding contradiction depends on the speed of adiabatic flux variation.}
    \label{fig:disc}
\end{figure}

Throughout this article, we used the adiabatic charge pump as a complementary tool. However, it was not applied in the definition of the procedure used to identify topological states. This is due to the fact that topology is not perfect for BLT states and many states will exhibit avoided crossings. Therefore, there might be discrepancies in terms of which states these methods deem to be topological. An example of such a possibility is seen in Fig.\ \ref{fig:disc} describing a BLT current in the system $(R_0, \phi, x_0, y_0)=(34, 0.7375, 17, 16)$. 

This could indicate that the procedure we defined is incomplete and needs amendments. One possible way is to create another procedure from the adiabatic charge pump, then demand topological states to be identified as such by both procedures. Building an procedure using the adiabatic charge pump can prove very cumbersome, as it needs to track all states throughout the flux variation. When tracking a state by an avoided crossing, one would have to take hybridization into account using fidelity calculations. This can be calculated as in Ref.\ \cite{Saswat} by using $\braket{\psi_{m}(\varphi(t))}{\psi_{n}(\varphi(t-dt))}$, which is the projection of the $n$th state at time $t-dt$ onto the $m$th at time $t$. This quantity depends on many factors such as the real-space localization of the states and the speed at which the flux is varied. Furthermore, the states pumped across the Fermi energy from below and the ones pumped across from above must be localized far enough apart in real-space in order to contribute to topology as seen in Eq.\ (\ref{eq:pump}). 

These factors make it difficult to determine which avoided crossings count as discrepancies with respect to our original procedure. For the system $(34, 0.7375, 0.1, 0.3)$ and $(34, 0.7375, 17, 16)$ possible discrepancies were rarely found and only at energy gaps of magnitude $10^{-4}$ or below as in Fig.\ \ref{fig:disc}.  

In defining our initial procedure, we used many subjective choices of thresholds and measures, which can be optimized through further testing. The measure for robustness $\Omega_{i}^Q$ was chosen in a way that allows efficient computations. Another method could be to compute $Q_{\text{dis}}(E_i)$ over a reasonable interval of disorders $W$ and record at which disorder $Q_{\text{dis}}(E_i)$ falls below a certain threshold. This is much more numerically taxing than our original definition as each disorder would have to be calculated numerous times to decrease uncertainty. 

For our procedure, we also made the choice of how to define a BLT state. This definition was a demand that the maximal cumulated crosshair marker lies close to a non-zero integer, and that the current is robust to a certain degree. However, one could choose to define the BLT states in other ways. One way is to demand that the cumulated crosshair marker $\Lambda(r, E)$ plateaus at an integer for a certain duration regardless of whether the maximal cumulated crosshair marker is close enough to an integer or not. For the case of the system $(34, 0.7375, 0.1, 0.3)$ we have $\abs{S}=914$. If we look for states where the cumulated crosshair marker plateaus within a deviation of $\epsilon=0.01$ over a radius difference of $\Delta r = 10$ then we get $106$ extra states to test. However, in many cases, the maximal value of the cumulated crosshair marker is not much higher than the plateaus, so most of these states are incorporated by our original procedure by loosening the requirements in the first sorting step in Eq.\ (\ref{eq:top_index}). Changing the requirement from $\epsilon=0.01$ to $\epsilon=0.03$ reduces the number of extra states from $106$ to $16$. These changes should not have any significant impact on our results and the properties found for BLT states. 

We found that some BLT states gather in groups while others are isolated. As such, further restrictions on topology can be imposed, demanding that any identified BLT state must be part of a group in order to secure its robustness against other kinds of disorder. 

\section{Conclusion}\label{sec:conclusion}

In this paper, we have created a procedure used to identify BLT states and edge states in quasicrystals. The procedure is based on sorting states using the maximal value of the cumulated crosshair marker followed by a subsequent sorting step where we defined and applied a topological measure $\Omega^Q_i$ and a radial measure $\Omega^R_i$. We then defined a method for identifying edge states with support on the whole perimeter of the graph and verified that the number of these edge states corresponds to the number of edge states identified by the procedure. We then applied the procedure to compute the number of BLT states and edge states for varying system sizes and magnetic fluxes. From this we found that the number of BLT states spikes for $\phi \in (0.65, 0.9)$ while being substantially smaller outside of this interval. The number of topological edge states stayed approximately constant for all fluxes in accordance with expectation except for the case where higher Chern numbers were found. This is because higher Chern numbers result in boundary currents penetrating further into the material. For varying radius, we found that the number of BLT states follows a linear trend in the number of sites with a slope of $0.098$. 

We then looked into the density of states for the topological states, showing that BLT states can occur at most values, unlike edge states. The BLT states were found to follow certain patterns but could also have disordered shapes. A large number of the BLT states were found to have circular shapes following the pattern drawn by sites with a coordination number of seven. Another prevalent pattern was that of smaller circular shapes each centered around points of coordination number eight. We found that at higher density of states most BLT states would follow these kinds of patterns, which indicates that the geometric support of the currents could induce greater robustness. We found that BLT states could occur both isolated and in groups, while adjacent non-topological states resemble them to a large degree. This could imply that BLT states must appear in groups to ensure complete robustness against different kinds of disorder. Furthermore, we found that the spatial localization of the BLT current could change throughout a given group of BLT states.  

Finally, we discussed possible discrepancies between our procedure and the results of the adiabatic charge pump. These cases were rarely found and only for band gaps of magnitude $10^{-4}$ or smaller. Because of this along with the fact that jumping an avoided crossing also depends on how close we are to the adiabatic limit, the presence of possible discrepancies is not immediately clear. Nevertheless, we discussed another procedure based on the adiabatic charge pump that could be used to refine the identification process. We also discussed possible redefinitions of our original procedure based on how one might choose to define a BLT state. This includes considering any case where the cumulated crosshair marker plateaus at a non-trivial integer independent of where the maximal cumulated crosshair lies. Furthermore, possible BLT states could be discarded depending on whether they are isolated or in a group of BLT states to ensure robustness.  

For further developments, one could work with optimizing the procedure with respect to the points we discussed in Sec.\ \ref{sec:Discuss} and see how this changes the results. The procedure we created can be applied to other quasicrystals and other Hamiltonians in order to study which topological properties they might exhibit, and whether certain patterns emerge for these cases. This could give a deeper insight into what properties dictate the existence and robustness of these kinds of states. As for the AB-tiling, one can look further into the properties of the identified BLT states. This entails numerically simulating the charge transport through time evolution and classifying the properties of the states depending on their patterns. Furthermore, one could study why the BLT states predominantly emerge only at a specific interval of magnetic fluxes.  

\begin{acknowledgements}
    We thank Saswat Sarangi and Callum W. Duncan for discussions. The work presented in this article is supported by Novo Nordisk Foundation grant NNF23OC0086670
\end{acknowledgements}

\appendix
\section{Chiral Symmetry}\label{appendix:sym}

In order to check for chiral symmetry of the model, we need to check whether our graph is bipartite. This means that the vertices of the graph can be sorted into two subsets, A and B, such that there are no edges between vertices belonging to the same sub-graph. To show this property for the AB-tiling we can prove a slightly more general statement: 

\textit{Any finite and two-dimensional plane graph is bipartite if and only if each tile of the graph has an even number of edges.} 

\textit{Proof:} We start by proving the simplest direction. Suppose we have a bipartite graph, and suppose towards contradiction that it contains a tile with an odd number of edges. Then it is impossible to assign each vertex of the tile to subgraphs A and B such that no vertices of the same subgraph are connected. This leads to a contradiction in our original assumption, meaning our graph only contains tiles with an even number of edges. 

We now prove the other direction. To prove that any graph consisting of tiles with an even number of edges must be bipartite, we apply mathematical induction in the number of tiles, $n$. The case $n=1$ is quickly seen to be true simply by assigning vertices along the edge alternately to A and B. \smallskip

Now, assume towards induction that our statement is true for graphs consisting of $n-1$ tiles, and assume we have an arbitrary graph consisting of $n$ tiles. Removing any of the tiles along the edge of the graph yields another graph which by our induction assumption must be bipartite. Then the edge of the graph must in particular also be bipartite, and so is the tile we removed. We can therefore assign the vertices on the graph-edge and on the tile to subgraphs A and B in a compatible way such that adding the removed tile to our graph again only overlaps vertices belonging to the same subgraph. This yields our original graph of $n$ tiles which by construction must be bipartite. 

This proof only works for finite-sized graphs, but in this paper we only deal with systems with open boundary conditions, so the above statement suffices. Since the AB-tiling consists only of four-sided tiles, it must be bipartite. Let us therefore denote sites belonging to each subgraph as $\textbf{r}_i^A$ and $\textbf{r}_i^B$ respectively. From this we can define a chiral symmetry operator, $\Gamma$, by
\begin{equation}
    \Gamma \ket{\textbf{r}_i^A} = +1 \ket{\textbf{r}_i^A}, \qquad \Gamma \ket{\textbf{r}_i^B} = -1 \ket{\textbf{r}_i^B}.
\end{equation}

One can readily check that the operator is both unitary and Hermitian, $\Gamma \Gamma^\dagger = \Gamma^2 = \mathbb{I}$ and satisfies $\Gamma H \Gamma^\dagger = -H$.

\section{Implications of Chiral Symmetry}\label{appendix:cross}

The existence of chiral symmetry for our system has different consequences. The most noticeable implication is that the spectrum of the Hamiltonian is symmetric, see Fig.\ \ref{fig:energy_spectrum}, as every eigenstate $\ket{\psi}$ with energy $E$ has a chiral symmetric partner $\Gamma \ket{\psi}$ with energy $-E$. This is quickly seen from $H\Gamma \ket{\psi} = -\Gamma H \ket{\psi} = - E \Gamma \ket{\psi}$. Another consequence of chiral symmetry is the relation given by
\begin{equation}
    C_{\textbf{R}}(\textbf{r}, -E) = -C_{\textbf{R}}(\textbf{r}, E).
\end{equation}

To prove this equation, we first rewrite the crosshair marker from Eq.\ (\ref{eq: crosshair}) to the form
\begin{align}
     C_{\textbf{R}}(\textbf{r}, E) = & -2\pi i \bra{\textbf{r}} P(E)\vartheta_{R_x}P(E)\vartheta_{R_y}P(E)
     \ket{\textbf{r}}  \\
    & + 2\pi i \bra{\textbf{r}} P(E)\vartheta_{R_y}P(E)\vartheta_{R_x}P(E)\ket{\textbf{r}} \notag
\end{align}
where we used that $\Tr_{\textbf{r}}{(...)}=\bra{\textbf{r}}...\ket{\textbf{r}}$. We can then relate the projection operator $P(-E)$ to $P(E)$ using chiral symmetry by
\begin{equation}
    P(-E) = \mathbb{I}-\Gamma P(E) \Gamma^\dagger.
\end{equation}

This can be inserted into the expression for the crosshair marker at energy $-E$ yielding the expectation value of sixteen terms. One can use the fact that $\vartheta_{R_x}, \vartheta_{R_y}$, and $\Gamma$ all commute among themselves, and that $\ket{\textbf{r}}$ is an eigenvector of each of these operators. Furthermore, denote the eigenvalue of $\Gamma$ as $\alpha \in \{\pm 1\}$ such that $\Gamma \ket{\textbf{r}} = \alpha \ket{\textbf{r}}$. This leads to almost all the terms canceling pairwise leaving only
\begin{align}
     C_{\textbf{R}}(\textbf{r}, & -E) \notag \\
     =
     & 2\pi i \bra{\textbf{r}} 
     \Gamma P(E) \Gamma^\dagger \vartheta_{R_x} \Gamma P(E) \Gamma^\dagger \vartheta_{R_y} \Gamma P(E) \Gamma^\dagger \ket{\textbf{r}} \notag \\
    & -  2\pi i \bra{\textbf{r}} \Gamma P(E) \Gamma^\dagger \vartheta_{R_y} \Gamma P(E) \Gamma^\dagger \vartheta_{R_x} \Gamma P(E) \Gamma^\dagger \ket{\textbf{r}} \notag \\  
     = & 2\pi \alpha^2 i \bra{\textbf{r}} P(E)\vartheta_{R_x}P(E)\vartheta_{R_y}P(E)
     \ket{\textbf{r}} \notag \\
    & - 2\pi \alpha^2 i \bra{\textbf{r}} P(E)\vartheta_{R_y}P(E)\vartheta_{R_x}P(E)\ket{\textbf{r}} \notag \\ 
    = & - C_{\textbf{R}}(\textbf{r}, E),
\end{align}
where we used the unitary and Hermitian properties of $\Gamma$. This yields the desired expression. 

\nocite{*}

\bibliography{apssamp}

\end{document}